\documentclass[aip,jcp,amsmath,amssymb,reprint,numeric]{revtex4-1}
\usepackage{graphicx}
\usepackage{subfigure}
\usepackage{color}
\usepackage{amsmath}
\usepackage{bm}

\begin{document}
\title {Fickian Yet Non-Gaussian Behaviour: A Dominant Role of the Intermittent Dynamics}
\author{Sayantan Acharya}
\author{Ujjwal Kumar Nandi}
\author{Sarika Maitra Bhattacharyya}
\email{mb.sarika@ncl.res.in}
\affiliation{\textit{Polymer Science and Engineering Division, CSIR-National Chemical Laboratory, Pune-411008, India}}

\date{\today}


\begin{abstract}
We present a study of the dynamics of small solute particles in a solvent medium where the solute is much smaller in
size, mimicking the diffusion of small particles in crowded environment. The solute exhibits Fickian diffusion arising from non-Gaussian
van Hove correlation function. Our study shows that there are at least two possible origins of this non-Gaussian behaviour.
The decoupling of the solute-solvent dynamics and the intermittency in the solute motion, the latter playing a dominant role. In the former scenario when averaged
over time long enough to explore different solvent environments the dynamics recovers the Gaussian nature. In case of intermittent
dynamics the non-Gaussianity remains even after long averaging and the Gaussian behaviour is obtained at a much longer time.
Our study further shows that only for intermediate attractive solute-solvent interaction the dynamics of the solute is 
intermittent. The intermittency disappears for weaker or stronger attractions. 
\end{abstract}
\maketitle

\clearpage

\section{Introduction}

Diffusion of smaller particles in a crowded medium is ubiquitous in nature and has huge industrial and academic 
relevance \cite{granick_pnas,granick_nature,yethiraj,yasho,sebastian,slater,maroncelli,metzler,weitz,bagchi,sayantan,solar,evans}.
The diffusion of sodium chloride through granular soil bed is important in understanding the ground 
water contamination from solid waste landfills \cite{soil}.  In the study of conductance in polyelectrolytes the
knowledge of salt diffusion in polar polymer matrix is important. In metallurgy we need to have the knowledge of  
diffusion of alloying elements like
hydrogenated titanium powder into metal matrices like Titanium \cite{metal}. The diffusion 
of protein in cytoplasm, colloidal beads through dense actin filament network, mimicking the cytoskeleton 
\cite{weitz,granick_pnas}, skin care products through membranes can also be modeled in terms of small particle 
diffusion in crowded medium. The diffusion of the solute particle depends on its size, its interaction with the 
solvent and also the dynamics and the structure of the solvent \cite{sayantan}. When the mean square displacement is linear with 
time then the solute is expected to follow the laws of Brownian motion originally derived by Einstein to explain 
the diffusion of a large solute molecule through a medium \cite{ein-1,ein-2,ein-3}.

 According to the theory of classical random walk a system 
which shows Brownian motion/Fickian diffusion should also have Gaussian distribution of the displacement probability \cite{ein-2}. However, there have 
been a number of studies on a diverse range of systems showing the Fickian but non-Gaussian behaviour 
\cite{maroncelli,granick_pnas,granick_nature,yethiraj,sebastian,slater}.  Usually this phenomena is attributed to the slow 
changing environment compared to the timescale of motion of the small diffusing particle (solute) \cite{granick_nature,yethiraj,sebastian,slater}. It has 
been argued that this can lead to changing diffusivity of the solute as the solvent environment changes and has been
termed as diffusing diffusivity \cite{slater,sebastian}. The diffusing diffusivity leads to Fickian but non Gaussian behaviour at short times,
however the displacement distribution is shown to become Gaussian at longer times \cite{slater,yethiraj,sebastian,granick_nature}.
Note that in supercooled liquids, although there is no timescale difference between the solute and the solvent dynamics, a similar
observation of Fickian but non-Gaussian distribution of displacement probability  has also been observed. In supercooled liquids and granular 
medium domains of different dynamical properties (fast and slow) form and the diffusion is found to 
become intermittent due to the formation of these domains \cite{pinaki}. Chaudhuri $et$ $al$ have shown that the dynamic 
heterogeneity which leads to the decoupling of the diffusion and the structural dynamics is responsible for the 
tail in the distribution function. There has also been studies where this anomaly in the dynamics has been attributed 
to weak ergodicity breaking \cite{metzler}.

Thus we find that this Fickian but non-Gaussian behaviour of the dynamics is omnipresent in a wide range of systems. In this work we 
aim to explore the origin of this non-Gaussian behaviour in details via computer simulation studies. We study
 diffusion of small solute particles through solvent molecules where the solvent molecules are always bigger in size compared to the 
solute. We vary the mass of the solvent and also the solute-solvent interaction energy from 
repulsive to strongly attractive. We find that in most of the cases the solute dynamics are Fickian but non-Gaussian. 
Our study reveals that  
there are two possible origins of non-Gaussian distribution of displacement probability, one arising from decoupling of 
the solute from the solvent
dynamics and the other from intermittency in the solute motion.
For systems which show only decoupling in dynamics,
when averaged over a long time the probability distributions recover 
their Gaussian nature but for systems where the solute dynamics is also intermittent the displacement probability distribution
remains non-Gaussian, even at long times. The intermittency seems to provide dominant contribution in the non-Fickian dynamics.

The next section contains simulation details. Section 3 contains the results and 
discussion followed by the conclusion in section 4.

\section{Simulation Details}
\label{sim}
In this work we perform an equilibrium Molecular Dynamics (MD) simulation with an 
atomistic model where particle type \textquoteleft i' is interacting with particle 
type \textquoteleft j\textquoteright  with 
truncated and shifted Lennard-Jones (LJ) pair potentials, given by
\begin{equation}
 \Phi_{ij}(r_{ij})={\Phi{_{ij}}^{LJ}(r_i,\sigma_{ij},\epsilon_{ij})-\Phi{_{ij}}^{LJ}(r{_{ij}}^{(c)},
\sigma_{ij},\epsilon_{ij})}, r_{ij}\leq{r{_{ij}}^c}
                \end{equation}
                 \label{2}
                 \begin{equation}
                  = 0 , r_{ij} > {r{_{ij}}^c}
\label{1}
\end{equation}
Where
 $\Phi{_{ij}}^{LJ}(r_{ij},\sigma_{ij},\epsilon_{ij})=4\epsilon_{ij}[{(\frac{\sigma_{ij}}{r_{ij}})}^{12}-
{(\frac{\sigma_{ij}}{r_{ij}})}^{6}]$,
$r_{ij}$ is the distance between pairs, $r{_{ij}}^{c}=2^{\frac{1}{6}}\sigma_{ij}$ for Weeks 
Chandler Anderson (WCA) system \cite{wca21} and $r{_{ij}}^c=2.5\sigma_{ij}$ for LJ system.
Here $i, j = 1, 2$, where 1 refers to solvent and 2 refers 
to solute. For all the systems
we consider that the  interaction between the solvent and the solute has a soft core
 which allows inter-penetration between solute-solvent pair.
 
 We take a system with 1000 particles where 10 of them are solutes and rest are solvents. 
In this simulation we study different systems varying 
the interaction potential (both attractive and repulsive), 
solvent mass and the interspecies interaction length.
 
 We have done isothermal-isobaric ensemble (NPT) simulation at a reduced temperature 1.663 (initially the 
crystals are melted in a higher temperature, later it is cooled down) and at reduced pressure 7 where the average density 
of the system is 0.8867 which ensures a crowded environment.
 
 The MD simulations are performed in a cubic box  using Nos\'{e}-Hoover thermostat and barostat \cite{nose,hoover}. The integration 
step is 0.001$\tau$, where $\tau=\sqrt{\frac{M\sigma_{11}^2}{\varepsilon}}$, where the solute mass $M=1$.
 In this study, length and temperature 
are given in the units of , 
$\sigma_{11}$ and $\frac{k_{B}T}{\varepsilon}$. We consider solvent mass, '$m$', over a wide range, 
0.5,1,10 and 100 for both attractive and repulsive systems. $\sigma_{11}$ is taken as 1, $\sigma_{22}$
 is .171 and $\sigma_{12}=\sigma_{22}+.171$. All the above mentioned systems are equilibrated for 1-2 ns followed by a production 
run of 4 ns (in Argon units). Systems with larger mass are equilibrated over longer times.

 We perform Molecular Dynamics (MD) simulations using LAMMPS  package \cite{lammps}.

We calculate the relaxation times from the decay of the self part of the overlap function $q(t)$ \cite{shila}. This is a
two point time correlation function of local density $\rho(r)$ and is 
defined as-
\begin{equation}
\langle{q(t)}\rangle\approx\langle\sum_{i=1}^{N}\delta(r_{i}(t_{0})-r_{i}(t+t_{0}))\rangle.
\label{2}
\end{equation}
\noindent
Here, $r$ is the position of the particle. The function $\delta$ is  approximated by a Heaviside step
function $\Theta(x)$. It defines the condition of overlap between two particle positions separated by a time interval t,
\begin{equation}
\langle{q(t)}\rangle\approx\langle\sum_{i=1}^{N}\Theta(|r_{i}(t_{0})-r_{i}(t+t_{0})|)\rangle,
\label{3}
\end{equation}
                  $\Theta(x)=1$, when $x\leq$ a, implying overlap, otherwise $\Theta(x)=0$.

However, this time dependent overlap function depends on the
choice of the cut-off parameter a, which we  take as 0.3.
The choice of this parameter is such that
particle positions separated by the  small amplitude vibrational motion are treated as the same, or that $a^{2}$
is comparable to the value of the MSD in the
plateau between the ballistic and diffusive regimes.

\begin{table}
 \caption{Solute, solvent diffusion values $D_1$ and $D_2$ respectively and their ratio $D_2/D_1$ for different systems. The solvent $\tau_{\alpha}$ values are also mentioned for every system.
In every system $\epsilon_{11}=$1, $\epsilon_{22}=$0.5 and $\epsilon_{12}=$6.}
\begin{center}
    \begin{tabular}{ | l | l | l | l | l | p{2cm} |}
    \hline
    system & solvent mass (m) & $\tau_{\alpha}$ & $D_1$ & $D_2$ & $D_2/D_1$ \\ \hline
    LJ & 0.5 & 0.25 & $0.119\pm .001$& .5 & $4.20\pm .04$ \\ \hline
    LJ & 1  & 0.357 &  $0.08\pm .0004$ & .35 & $4.37\pm .02$  \\ \hline
    LJ & 10  & 1.08 & $0.0278\pm .0002$ & .23 & $8.27\pm .06$  \\ \hline
    LJ & 100 & 3.52 & $0.0088\pm .0002$ & .207 & $23.5\pm .5$ \\ \hline

    \end{tabular}
\end{center}
\end{table}
\noindent


\begin{table}
	\caption{Solute, solvent diffusion values $D_1$ and $D_2$ respectively and their ratio $D_2/D_1$ for different systems. The solvent $\tau_{\alpha}$ values are also mentioned for every system.
		In every system $\epsilon_{11}$ is 1 and $\epsilon_{22}$ is 0.5. Solvent mass m=100 here.}
	\begin{center}
		\begin{tabular}{ | l | l | l | l | l | p{2cm} |}
			\hline
			system  & $\epsilon_{12}$ & $\tau_{\alpha}$ & $D_1$ & $D_2$ & $D_2/D_1$ \\ \hline
			WCA & 1 & 3.43 & $0.0087\pm .0002$ & 1.6 &  $185\pm 5$ \\ \hline
			LJ  & 2 & 3.60 & $0.0087\pm .0002$ & 1.5 & $173\pm 5$ \\ \hline
			LJ  & 4 & 3.62 & $0.0084\pm .0001 $ & .44 & $52\pm 1$ \\ \hline
			LJ  & 6 & 3.53 & $0.0088\pm .0002$ & .207 & $23.5\pm .5$ \\ \hline
		    LJ  & 16 & 3.62 & $0.0084\pm .00004$ & .02 & $2.37\pm .01$ \\ \hline
			LJ  & 24 & 3.48 & $0.0087\pm .00004$ & .016 & $1.82\pm .01$ \\ \hline
		\end{tabular}
	\end{center}
\end{table}
\noindent

\section{Results and Discussion}

\begin{figure}[h]
	\centering
	\includegraphics[width=0.45\textwidth]{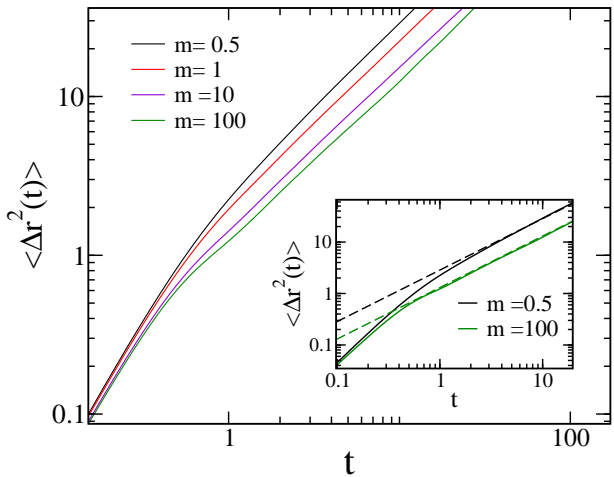}
	\caption{ \it{ $<\bigtriangleup r^{2}(t)>$ for the solute calculated from Eq. \ref{4} is plotted against time for attractive LJ potential with solvent mass= 0.5,1,10 and 100.
				All of them show clear linear regime 
				predicting diffusion dynamics.  (Inset) MSD plots for mass 0.5 and 100. 
				Linear region sets in at an earlier time for higher mass.
		}}
		\label{fig1}
	\end{figure}
	\noindent

	\begin{figure*}[h]
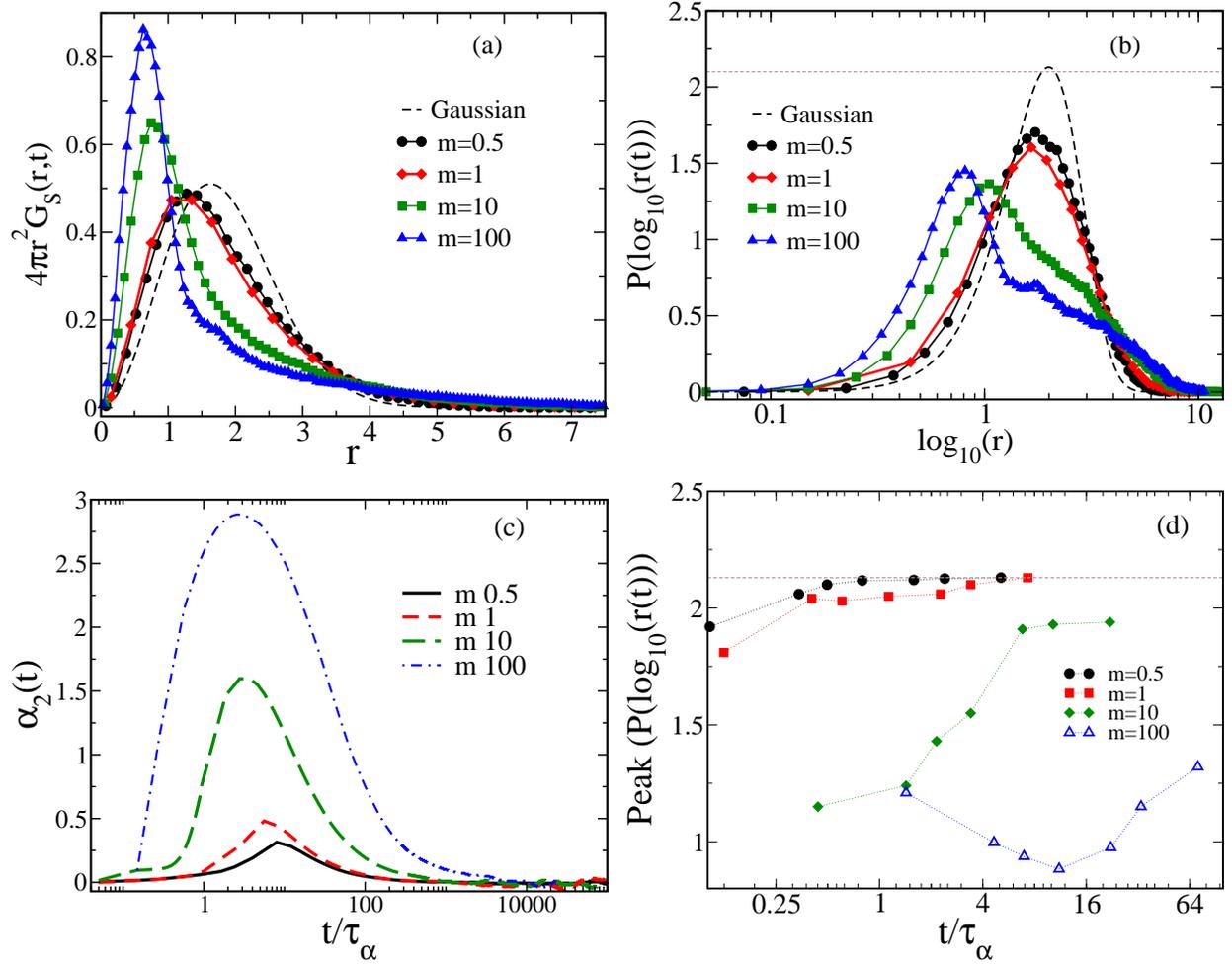

		\centering
		\subfigure{
			\includegraphics[width=0.45\textwidth]{fig2a.eps}}
		\subfigure{
			\includegraphics[width=0.45\textwidth]{fig2b.eps}}
		\subfigure{
			\includegraphics[width=0.45\textwidth]{fig2c.eps}}
		\subfigure{
			\includegraphics[width=0.45\textwidth]{fig2d.eps}}
		\caption{\it{ (a) $4\pi r^{2}G_{s}(r,t)$ vs. r for the solute in systems with LJ interaction ($\epsilon_{12}$=6) and different solvent mass, calculated at time
				where $<\bigtriangleup r^{2}(t)>\approx$4.
				For comparison we also plot the Gaussian behaviour, $G_{Gaus}(r,t)=[\frac{3}{2\pi \langle r^{2}(t)\rangle}]^\frac{3}{2}\exp{[-\frac{3r^{2}}{2\langle r^{2}(t)\rangle}]}$, 
				where  r(t) is the position of the particle at time t assuming that r(t=0)=0.
				(b) Corresponding
				$P(log_{10}(r(t)))$ vs. $log_{10}(r)$ plots. 
				For comparison we also plot the  $P_{Gaus}(log_{10}(r,t))$. Both the plots predict that
				departure from Gaussianity is larger for higher solvent masses. 
				(c) Plots of non-Gaussian parameter $\alpha_{2}(t)$ against time scaled by individual $\tau_{\alpha}$ mentioned in Table 1. Non-Gaussian parameter is defined as $\alpha_{2}(t)=\frac{3\langle \delta r^{4}_{\alpha}(t) \rangle}{5{ \langle \delta r^{2}_{\alpha}(t) \rangle }^{2}}-1$. (d) Corresponding $P(log_{10}(r(t)))$ peaks vs. time plots for LJ systems changing solvent mass. Both (c) and (d) \textcolor{blue}{demonstrate} that
					departure from Gaussianity is larger for higher solvent mass. }}
		\label{fig2}
	\end{figure*}
	\noindent

In this study we work with a solute-solvent system where the solute size is chosen to be small so that the solute can 
explore the inter-solvent cage \cite{sayantan}. In earlier studies \cite{granick_nature,yethiraj} it has been shown that
for such small solute particles the non-Gaussian behaviour of the van Hove correlation
function is due to decoupling in solute-solvent dynamics.
In order to understand how slowing down of solvent dynamics, which leads to the decoupling,
effects  the solute motion, we study a set of systems where we vary the solvent mass. 
As observed earlier \cite{rajesh,sayantan} with the increase in solvent mass the 
decoupling between the solute and the solvent dynamics increases as seen from the $D_2/D_1$ values in Table 1.
 In these studies the solute-solvent interaction is kept moderately attractive ($\epsilon_{12}$=6).
The mean square displacement (MSD) is calculated from single particle trajectory, r(t), and then time
averaging and also particle averaging is performed,

\begin{equation}
\overline{\bigtriangleup r^{2}(t)}=\frac{1}{N}\sum_{i=1}^{N}\frac{1}{T}\int_{0}^{T}[r(t+t')-r(t')]^{2}dt'.
\label{4} 
\end{equation}
\noindent

Here, T is the overall measurement time, N is the number of particle, which is 10 in our case.
\textcolor{blue}{At longer time the MSD plots of the solute in all the systems
are linear with time} and for higher solvent mass the linear region sets in at an earlier time  as has 
 been observed in previous studies \cite{yethiraj} (Fig.{\ref{fig1}}).  A possible reason for this observation can be,
 for large solute-solvent decoupling solute does not expect the solvent to have any dynamics thus the diffusive dynamics
 for the solute in these systems sets in earlier assuming the solvent to be static. 
The linearity of the MSD predicts that the motion is diffusive.

 The  van Hove correlation function which is the distribution of probability displacement ($G_{s}(r,t)=\frac{1}{N}\langle\sum_{i=1}^{N}\delta(r+r_{i}(0)-r_{i}(t))\rangle$), is
calculated at the time where the MSD 
is linear and $<\bigtriangleup r^{2}(t)>$=4. Although the MSD is linear the $G_{s}(r,t)$ shows non-Gaussian nature (Fig.\ref{fig2}).

\begin{figure}[h]
\centering
\includegraphics[width=0.45\textwidth]{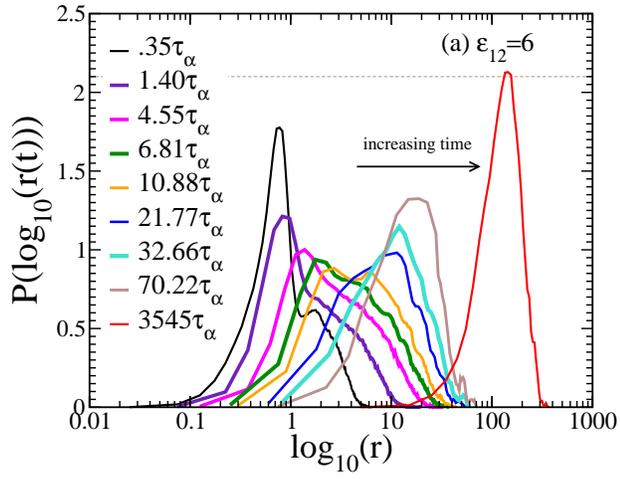}

\caption{ \it{The time evolution of $P(log_{10}(r(t)))$ vs. $log_{10}(r)$ for the solute  for LJ system ($\epsilon_{12}$=6) where the solvent mass is 100.
A clear bimodal nature in the solute dynamics indicates the presence of intermittency. Solute particles become Gaussian at much longer time $\sim 3500\tau_{\alpha}$.  The dashed line at 2.1 represents the value of the peak of a Gaussian function.}}
\label{fig4}
\end{figure}
\noindent

\begin{figure*}[h]
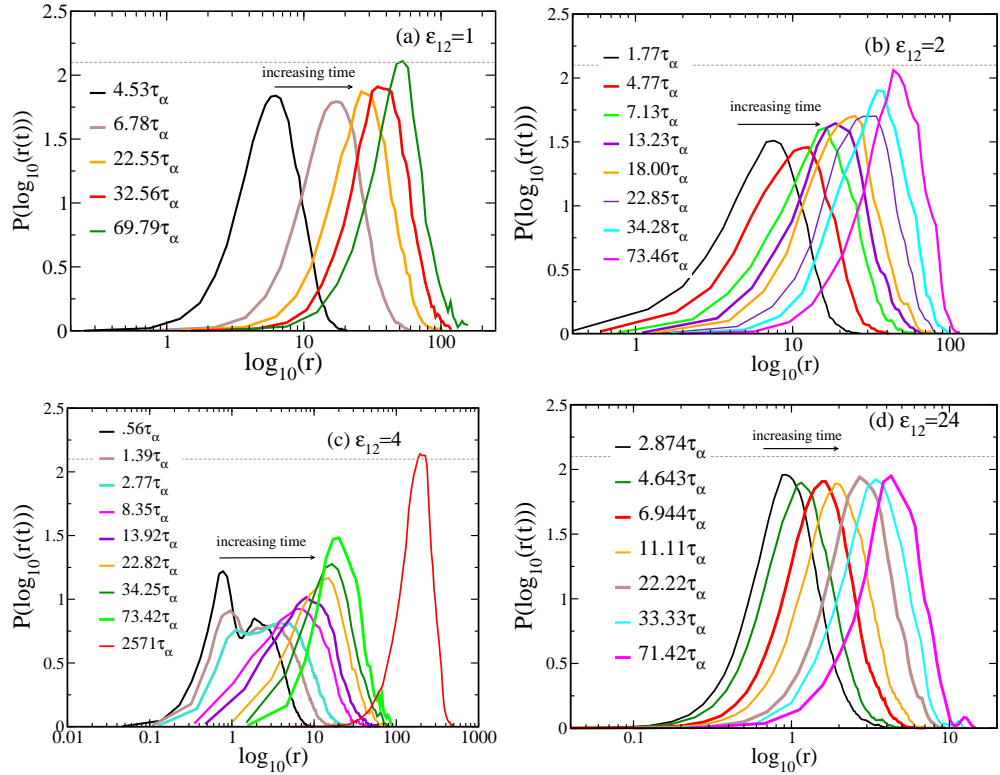

\centering
\subfigure{
	\includegraphics[width=0.36\textwidth]{fig4a.eps}}
\subfigure{	
 \includegraphics[width=0.36\textwidth]{fig4b.eps}}
\subfigure{
 \includegraphics[width=0.36\textwidth]{fig4c.eps}}
 \subfigure{
 \includegraphics[width=0.36\textwidth]{fig4d.eps}}

\caption{\it{ The time evolution of $P(log_{10}(r(t)))$ vs. $log_{10}(r)$ for solutes where the solvent mass is 100. (a) WCA system with $\epsilon_{12}$=1. System reaches Gaussianity at an earlier time and no bimodal signature is observed
indicating there is no intermittency in the dynamics. (b) LJ $\epsilon_{12}$=2.  At short
time it shows non-Gaussian behaviour and at longer time the behaviour is Gaussian. (c) Same as in (b) but for $\epsilon_{12}$=4.
At short and intermediate times the $P(log_{10}(r(t)))$ shows non-Gaussian and bimodal behaviour. (d) Same as in (b) for $\epsilon_{12}$=24. As compared to the other systems this system reaches Gaussianity at an earlier time. The dashed line at 2.1 in every plot represents the value of the peak of a Gaussian function.}}
\label{al6}
\end{figure*}
\noindent

\begin{figure}[h]
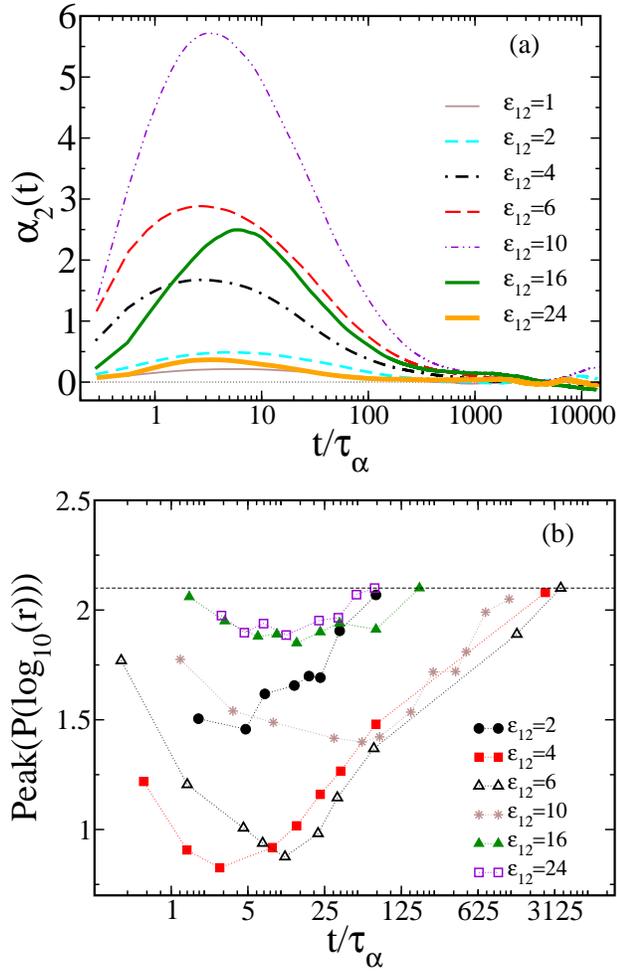

	\centering
	\subfigure{
		\includegraphics[width=0.45\textwidth]{fig5a.eps}}
	\subfigure{
		\includegraphics[width=0.45\textwidth]{fig5b.eps}}
	\caption{\it{(a) Plots of non-Gaussian parameter $\alpha_{2}(t)$ against t/$\tau_{\alpha}$ for LJ systems (m=100) having different $\epsilon_{12}$ values. $\tau_{\alpha}$ values are given in Table 2. Value of $\alpha_{2}(t)$ is maximum for  $\epsilon_{12}$=10. (b) Corresponding $P(log_{10}(r(t)))$ peaks vs. time plots.   }}
	\label{peak}
\end{figure}
\noindent

\begin{figure}[h]
	\centering
	\subfigure{
		\includegraphics[width=0.45\textwidth]{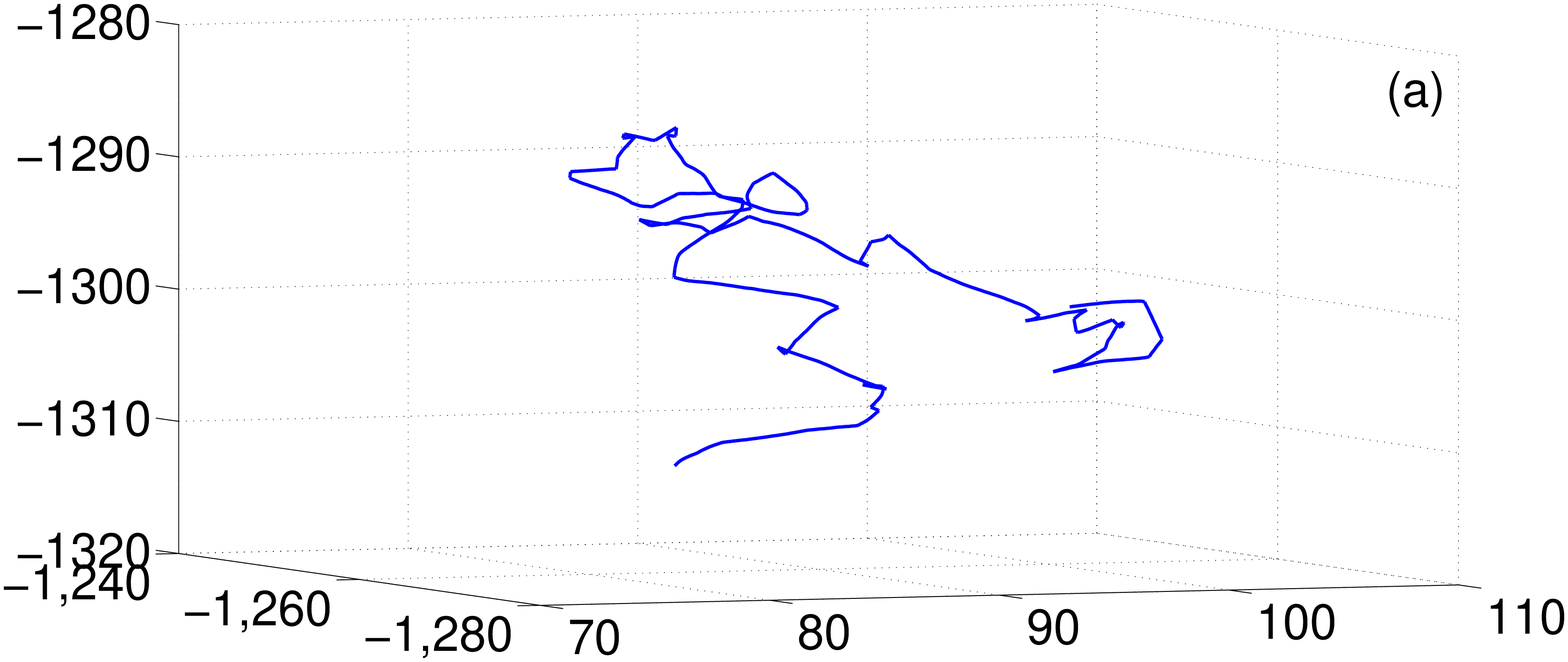}}
	\subfigure{
		\includegraphics[width=0.45\textwidth]{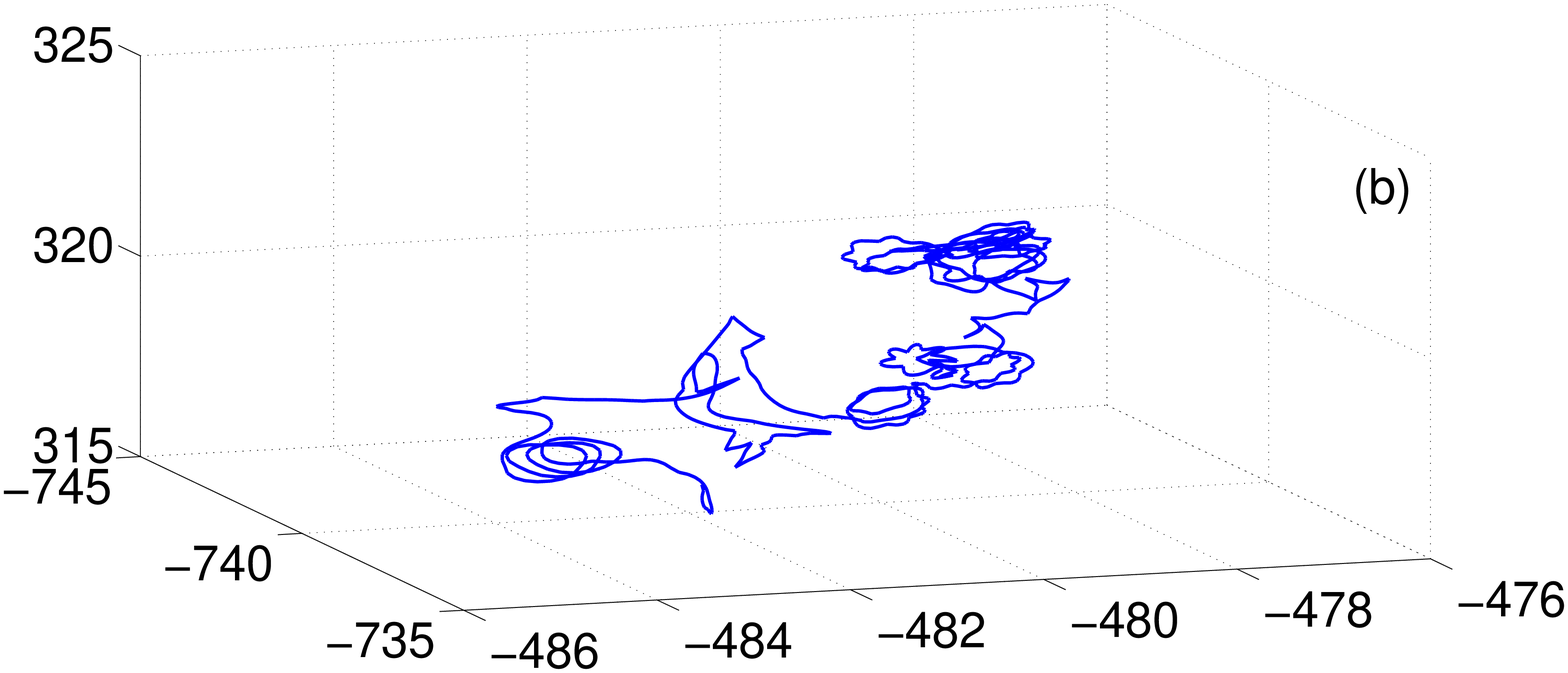}}
	\subfigure{
		\includegraphics[width=0.45\textwidth]{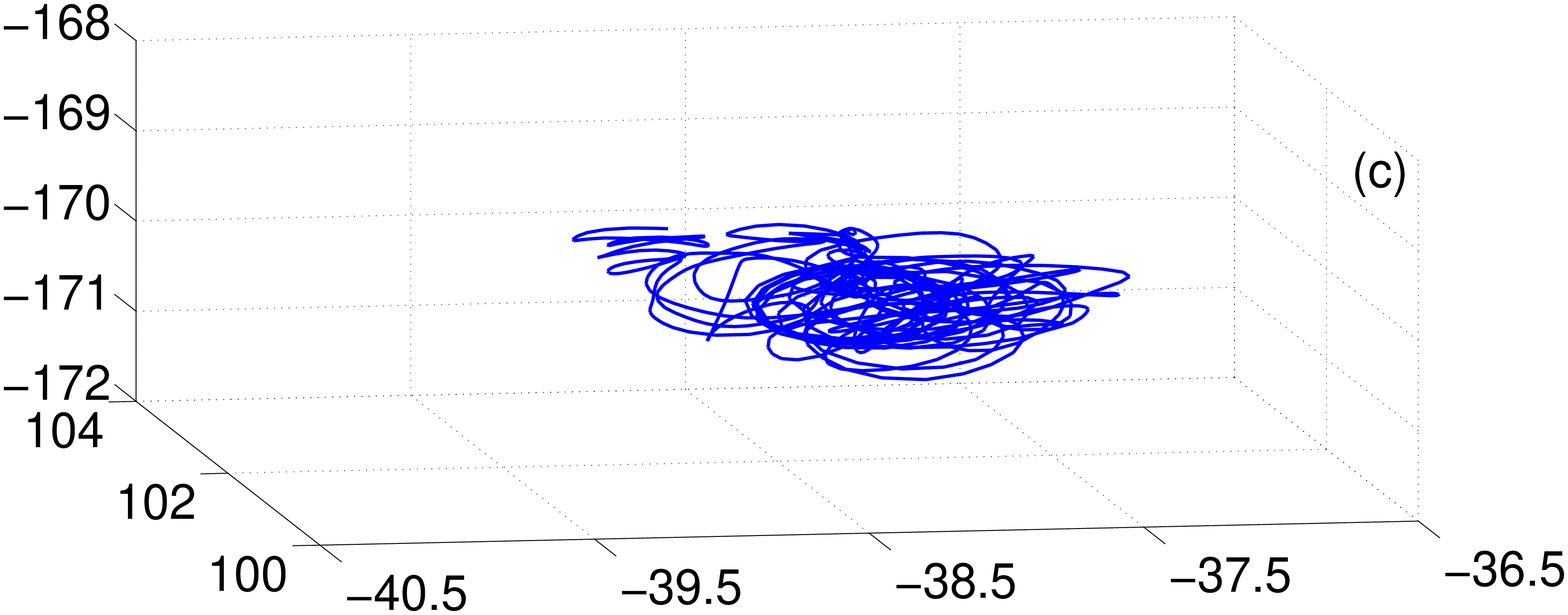}}
	\caption{\it{ Representative trajectory segments of 1000 time frames are shown for LJ systems when m=100. We have changed the solute-solvent interaction parameters and plotted them as (a) $\epsilon_{12}$=2, (b) $\epsilon_{12}$=4 and (c) $\epsilon_{12}$=24. We observe sticky points in case of  $\epsilon_{12}$=4 and 24.  In case of $\epsilon_{12}$=4 we also observe intermittency in the dynamics.}}
	\label{sticky}
\end{figure}
\noindent

\begin{figure}[h]
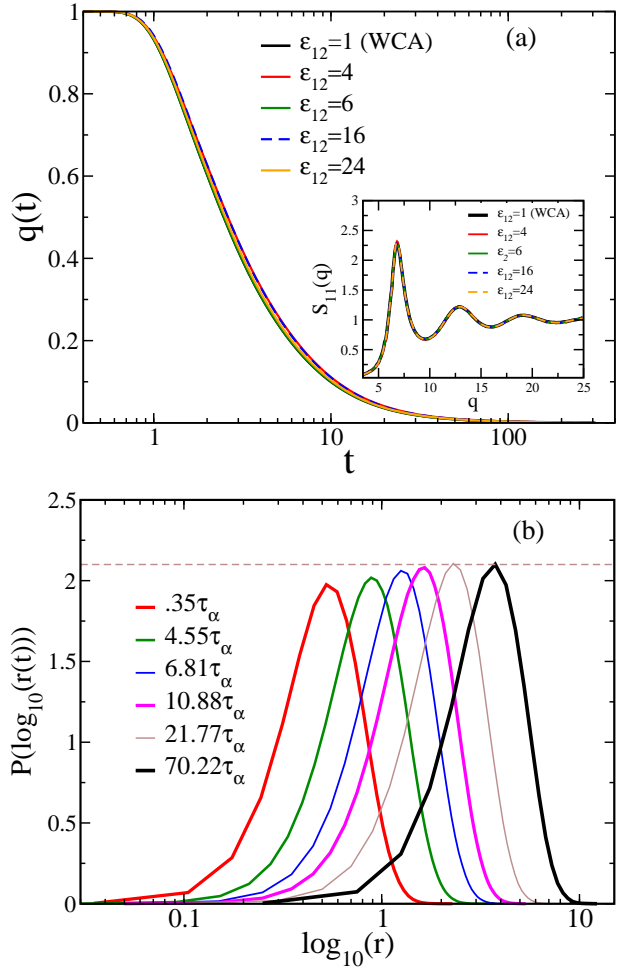

	\centering
	\subfigure{
		\includegraphics[width=0.45\textwidth]{fig7a.eps}}
	\subfigure{
		\includegraphics[width=0.45\textwidth]{fig7b.eps}}
	
	\caption{ \it{(a) Overlap functions q(t) vs. time for the solvent both for WCA and LJ systems ($\epsilon_{12}$=4,6,16,24) for m=100.
			We see there is hardly any difference indicating that the
			solvent dynamics for all attractive and repulsive systems are same. In the inset we show the corresponding
			solvent structures which are also similar. (b)The time evolution of $P(log_{10}(r(t)))$ vs. $log_{10}(r)$ for the solvent for LJ system ($\epsilon_{12}$=6) where m=100. Solvent reaches Gaussianity at an earlier time and no bimodal behaviour is observed.}}
	\label{sq}
\end{figure}
\noindent

As shown in Fig.\ref{fig2}a the non-Gaussian behaviour of $G_{s}(r,t)$ increases with the increase in solvent mass.
In order to measure the departure from Gaussianity we also plot $P(log_{10}(r(t)))= 4 \pi r^{3} ln (10) G_{s}(r,t)$ (Fig.\ref{fig2}b).
When the van Hove function is Gaussian then the peak value of $P(log_{10}(r(t))) \simeq 2.13$ \cite{szamel}. Its
decrease in value is a measure of the degree of departure from Gaussianity. 
 As expected, comparing figures Figs.\ref{fig2}a and \ref{fig2}b we find that systems which show larger departure from Gaussianity in
the former, has smaller peak values in the latter. 
We also plot the non-Gaussian parameter $\alpha_{2}(t)$ and the peak value of $P(log_{10}(r(t)))$ plots in Figs.\ref{fig2}c and \ref{fig2}d respectively. Both show that the departure from Gaussianity increases with mass.
 We find a positive correlation between the departure from Gaussianity and decoupling of solute-solvent dynamics which
is similar to what has been reported earlier \cite{granick_nature,yethiraj}. 
 It has been argued that although the solute reaches a 
diffusive dynamics, due to slow solvent dynamics it is unable to explore different solvent environments thus leading to
the non-Gaussian behavior of the probability distribution. It is found that for such systems the van Hove correlation
function at later times, which is long enough to explore the ergodic solvent dynamics, is Gaussian. Although our results apparently look
similar to that observed in earlier studies,
for our systems with large solvent mass we do not obtain the Gaussian behaviour even when the van Hove correlation 
function is calculated at about 70 times $\tau_{\alpha}$, where $\tau_{\alpha}$ is the $\alpha$ 
relaxation time of the solvent (Fig.\ref{fig4}).

Note that for attractive solute-solvent interaction and large solvent mass
we not only have the decoupling of the dynamics but the solute motion also becomes 
intermittent. The second peak or the shoulder in the $P(log_{10}(r(t)))$ plot is a signature of this intermittency which is 
observed for $m=$10 and 100 (Fig.\ref{fig2}(b)).

The intermittency in the solute motion appears quite similar to that found in supercooled liquids. In supercooled 
liquids there is formation of domains which give rise to the intermittency. In our present system there is no domain 
formation and the intermittency is the effect of attractive solute-solvent interaction.
While undergoing random walk, the solute explores the transient solvent cage \cite{sayantan,yasho}.
As the solute is small when it passes through the neck of the cage it feels uneven attraction from the different solvent particles and spends longer time
near a particular solvent, which can be envisaged as a sticky point and
leads to the intermittency in the solute dynamics.
Thus we may expect that larger the attraction the more intermittent is the dynamics.
In order to understand the 
connection between solute-solvent interaction, intermittency and non-Gaussianity of $G_{s}(r,t)$
we study a series of system by varying the strength 
of the solute-solvent interaction from repulsive to highly attractive. In all these cases the solvent
mass is kept at 100 where the LJ system ($\epsilon_{12}$=6) shows maximum deviation from  Gaussianity. 

For repulsive and small attractive interaction there is a large decoupling between the solute and the solvent dynamics
as seen from the $D_{2}/D_{1}$ values in Table 2. 
However the time evolution of $P(log_{10}(r(t)))$ does not show any bimodal nature thus predicting no strong
 intermittency in the solute dynamics (Figs.\ref{al6}a and \ref{al6}b). Thus, although the large solute-solvent decoupling initially 
gives rise to non-Gaussian $G_{s}(r,t)$, at longer times it becomes Gaussian as seen in Fig.\ref{al6}a for WCA system and Fig.\ref{al6}b for LJ system with $\epsilon_{12}$=2. On the other hand, for very large solute-solvent interaction
 the decoupling of the solute and the solvent dynamics reduces as seen from the
$D_2/D_1$ values in Table 2. Similar to the repulsive and small attractive systems, 
the time evolution of  $P(log_{10}(r(t)))$ does not show any bimodal nature thus the intermittency in the solute  
dynamics is also not present (Figs.\ref{al6}a and \ref{al6}b). Hence the dynamics remains almost Gaussian for short, intermediate and long times. However, for intermediate solute-solvent 
interaction ($2<\epsilon_{12}<16$) there is strong enough decoupling between the solute and the solvent dynamics
(Table 2) and strong intermittency in the 
dynamics as observed in the clear bimodal behaviour of the time evolution of the $P(log_{10}(r(t)))$ 
(Fig.\ref{fig4} and Fig.\ref{al6}c). This leads to persistent non-Gaussian dynamics even at longer times.
  Thus we find that although the solute-solvent decoupling decreases monotonically with increase in $\epsilon_{12}$ value, the intermittency and the non-Gaussianity in the dynamics has a non-monotonic dependence on the same. This non-monotonicity can also be observed in Fig.\ref{peak}a where we plot the non-Gaussian parameters for different $\epsilon_{12}$ values and in Fig.\ref{peak}b where we plot the peak values of $P(log_{10}(r(t)))$. Note that, both are measures of non-Gaussian behaviour and both show a non-monotonic dependence on $\epsilon_{12}$. However the $\epsilon_{12}$ values where we would expect the maximum departure from Gaussianity is different for the two parameters. The values of $\alpha_{2}(t)$ is maximum for $\epsilon_{12}$=10 whereas the departure of the peak value of $P(log_{10}(r(t)))$  w.r.t the value of 2.13, which is expected for a Gaussian function, is maximum for $\epsilon_{12}=$4 and 6. 

The change in the dynamics and also the existence of sticky points can also be observed by plotting the trajectories of the solute in different systems (Fig.\ref{sticky}). For $\epsilon_{12}$=2 we see almost a free movement of solute particle without any sticky points. For $\epsilon_{12}$=4 apart from free motion the trajectory looks coiled in certain region which indicates the presence of sticky points. Solute particles are almost stuck and move very little in case of $\epsilon_{12}$=24. Note the axis values of the figures are not same in the three plots.

In order to show that the solvent structure and dynamics have no role in the change of the solute dynamics with the change in $\epsilon_{12}$ value, we plot a few representative overlap functions of the solvent (Fig.\ref{sq}a) and also present the diffusion values in Table 2 for
$m=100$. We also show the corresponding solvent structure factors in the inset of Fig.\ref{sq}a. Note that the structure and dynamics of the solvent does not change with $\epsilon_{12}$, thus 
if the non-Gaussian behaviour is only due to the slow solvent dynamics then for all systems the recovery
of the Gaussian nature should happen over the same time period, which does not appear to be the case.  We have also claimed that unlike supercooled liquids the solvent remains homogeneous and intermittency in the solute dynamics is an effect of the intermediate solute-solvent attraction strength. To prove our point as a representative plot we show the solvent $P(log_{10}(r(t)))$ for $\epsilon_{12}$=6. The plot shows that when the solute is bimodal (Fig.\ref{fig4}) the solvent dynamics is Gaussian hence the latter does not play any role in the bimodal nature of the former (Fig.\ref{sq}b). 

Thus our study predicts that in crowded systems at least there are 
two different sources of non-Gaussian  behaviour. One is the  difference in timescale of the solute and solvent dynamics where the Gaussian nature is recovered at longer times,
and the other is the intermittency in the solute dynamics which arises due to attractive solute-solvent interaction and in this case non-Gaussian behaviour persists for a long time.
The intermittency in the solute dynamics
provides a dominant contribution to the non-Gaussian nature of $G_{s}(r,t)$.

\clearpage
\section{Conclusion}
When the dynamics of a particle is Brownian  it is supposed to have Gaussian distribution of its displacement probability (van Hove correlation function) which eventually gives rise to Fickian diffusion where the MSD remains linear with time. However there are systems where although the diffusion is Fickian the distribution of displacement probability is non-Gaussian. This anomaly has been explained as an effect of different phenomena like weak ergodicity breaking \cite{metzler}, decoupling of solute-solvent motion \cite{granick_nature,yethiraj} and diffusing diffusivity \cite{sebastian,slater}. In order to further understand this Fickian but non-Gaussian dynamics in this work we study a wide range of solute-solvent systems where the solute size is kept small so that it can explore the inter solvent cage and mimic the diffusion in a crowded environment. The mass of the solvent is varied to study the effect of the solute-solvent dynamical decoupling on the solute motion and the solute-solvent interaction is also varied to study the effect of it on the solute dynamics.

We find that similar to that reported earlier as the decoupling of the solute and the solvent dynamics increases the van Hove correlation function becomes more non-Gaussian \citep{granick_nature,yethiraj}. However, for repulsive and weak attractive values of solute-solvent interaction the dynamics becomes Gaussian when averaged over long enough time so that the solute explores all possible solvent environments. For these systems the origin of the non-Gaussian dynamics is the solute-solvent decoupling. We find that apart from this there is at least one more phenomena, the intermittency in the solute dynamics, which gives rise to non-Gaussian nature. Intermittency in particle dynamics, also giving rise to non-Gaussian dynamics, is usually reported in supercooled liquids \cite{pinaki}. The origin of this intermittency is the formation of dynamically heterogeneous domains. In our system there is no such domain formation and the intermittency is an effect of attractive interaction between the solute and the solvent. Surprisingly the intermittency is maximum for intermediate solute-solvent attraction
and reduces substantially when the solute-solvent attraction is decreased or increased. The reason behind this is to produce intermittency the attraction on one hand should be weak enough to have decoupling in the solute-solvent dynamics and on the other hand strong enough to produce sticky points in the solute path.  In case of intermittent dynamics the non-Gaussian behaviour persists over a longer time.

Thus our study reveals that for diffusion in a crowded environment for certain range of attractive interaction between the solute and the solvent it is possible to have intermittency in the dynamics and this can play a dominant role in giving rise to the non-Gaussian but Fickian dynamics.




\end{document}